\documentclass[%
twocolumn,
superscriptaddress,
nofootinbib,
longbibliography,
aps,
prl
]{revtex4-2}

\usepackage{amsmath,amssymb,bm,braket,url,dsfont,xspace,enumitem}
\usepackage{graphicx}
\usepackage{xcolor}
\usepackage{array}
\usepackage{booktabs}

\usepackage{longtable}

\usepackage[
draft=false,
bookmarks=true,
bookmarksnumbered=false,
bookmarksopen=false,
bookmarksopenlevel=4,
colorlinks=true,
anchorcolor=black,
citecolor=blue,
filecolor=magenta,
linkcolor=blue,
linkbordercolor={1 0 0},
urlcolor=blue,
pdfborder={0 0 1},
pdftitle={},
pdfauthor={},
pdfsubject={},
pdfkeywords={},
pdfpagemode=UseThumbs,
]{hyperref}

\usepackage{orcidlink}

\newcommand{\bk}{{\bm{k}}}

\newcommand{\T}{$\theta$\xspace}

\newcommand{\sca}{\mathcal{O}}

\begin{document}
\title{
Dual-Circular Raman Optical Activity of Axial Multipolar Order
}

\author{Hikaru Watanabe
        \orcidlink{0000-0001-7329-9638}
        }
\affiliation{Department of Physics, University of Tokyo, Tokyo 113-0033, Japan}
\affiliation{Department of Applied Physics, Hokkaido University, Sapporo 060-8628, Japan}

\author{Rikuto Oiwa
        \orcidlink{0000-0001-6429-6136}
        } 
\affiliation{Graduate School of Science, Hokkaido University, Sapporo 060-0810, Japan}
\affiliation{Center for Emergent Matter Science, RIKEN, Wako 351-0198, Japan}~~

\author{Hitoshi Mori
        \orcidlink{0000-0003-2901-8602}
        } 
\affiliation{Institute for Materials Research, Tohoku University, Sendai 980-8577, Japan}~~

\author{Ryotaro Arita
        \orcidlink{0000-0001-5725-072X}
        } 
\affiliation{Department of Physics, University of Tokyo, Tokyo 113-0033, Japan}
\affiliation{Center for Emergent Matter Science, RIKEN, Wako 351-0198, Japan}~~

\begin{abstract}
        Multipolar order, such as octupolar order, is a key concept in condensed matter physics, particularly in light of elusive hidden orders.
        However, its experimental identification remains challenging due to the absence of direct coupling to conventional external stimuli.
        In this study, we propose that dual-circular Raman scattering serves as a probe of multipolar anisotropies.
        By combining symmetry analysis with microscopic calculations, we identify that both time-reversal-even (\T{}-even) and time-reversal-odd (\T{}-odd) axial multipolar phases exhibit the sizable Raman optical activity as a direct consequence of multipolar symmetry breaking.
        The quantitative significance of the proposed response is demonstrated by the first-principles study of pyrite, a prototypical axial octupolar material.
        Furthermore, we reveal that a multipolar phonon, a three-dimensional and alternating displacement resembling the chiral phonon, plays a vital role in the proposed optical phenomena. 
        Our findings open a pathway for identifying multipolar orders in various materials through dual-circular Raman spectroscopy as a sensitive and versatile probe.
\end{abstract}

\maketitle

                \begin{figure}[h]
                \centering
                \includegraphics[width=\linewidth,clip]{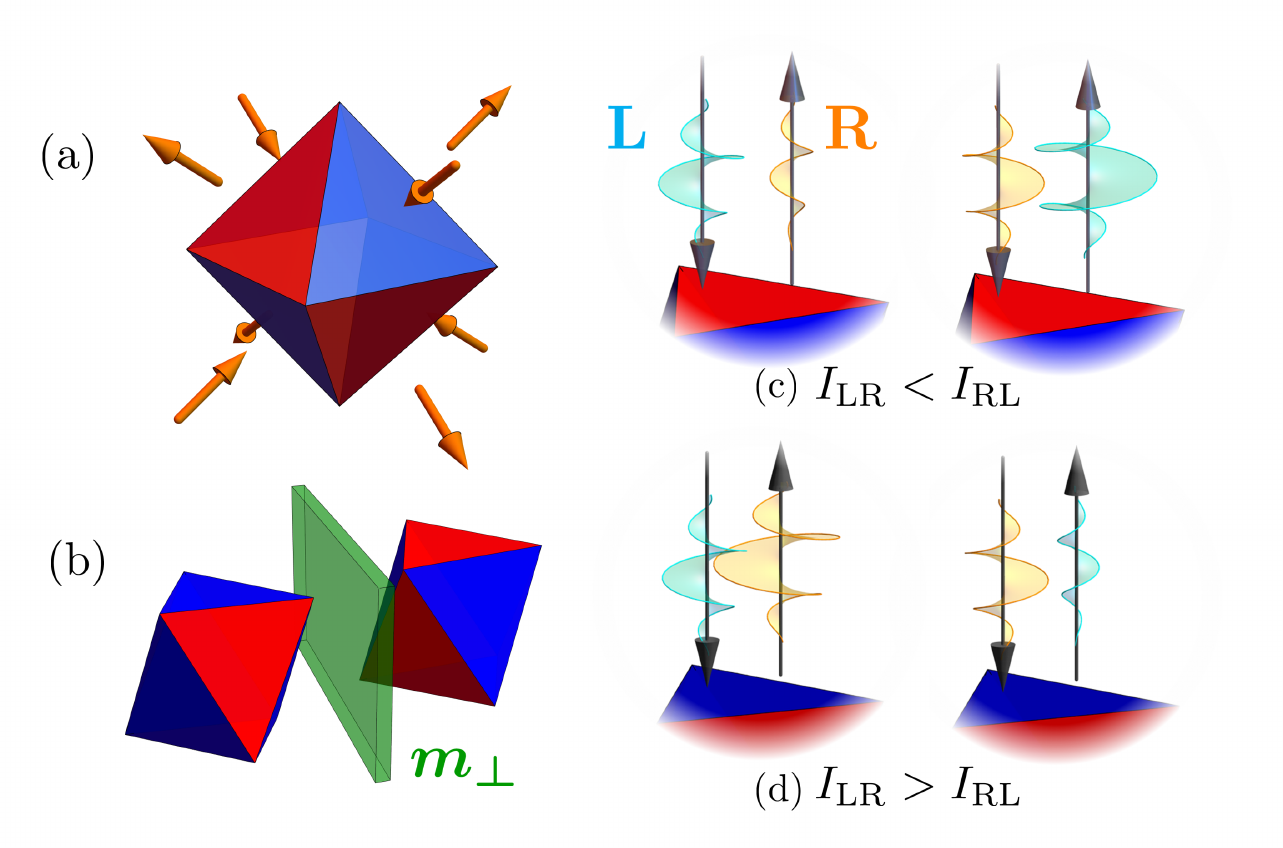}
                \caption{
                (a) $xyz$-type axial octupolar order.
                Faces of the octahedron are colored according to the penetrating direction of axial vectors (orange arrows) on each surface.
                (b) Mirror operation $m_\perp$ perpendicular to the octahedron's faces, illustrated by the green-colored mirror plane, flips the polarization of axial vectors, leading to the reversal of octupolarization.
                (c,d) Cross-circular Raman optical activity (ROA) of the axial multipolar systems.
                The conversion rate of the circular polarization of light differs between the right-to-left ($I_\text{RL}$) and left-to-right ($I_\text{LR}$) changes, depending on the axial polarization, which corresponds to the facets.
                Cross-circular ROA ($I_\text{RL} - I_\text{LR}$) for (c) the $\left\{ 111\right\}$ incidence and (d) the $\left\{ \bar{1}11\right\}$ incidence show the opposite signs.
                        }
                \label{Fig_1}
                \end{figure}

\noindent\textit{Introduction---}
Multipolar order is an ordered state exhibiting higher-order electromagnetic anisotropy, such as quadrupole and octupole.
It appears in systems where various interactions such as electronic correlation and spin–orbit coupling are significant, leading to intriguing physical phenomena~\cite{Kuramoto2009-sn,Santini2009-ex,Witczak-Krempa2014-zt,Onimaru2016-ea,Suzuki2018-me}.
Multipolar degree of freedom has recently been discussed in a wide range of contexts, including topological materials, quantum criticality, spin liquid, and spintronics~\cite{Wan2011-vp,Ueda2017-xk,Si2001-lw,Sakai2011-la,Bauer2002-ok,Ikeda2012-gu,Sibille2020-ow,Nakatsuji2015-iq,Suzuki2017-ps,Hayami2024-gd,jungwirth2025altermagneticspintronics}.

Despite its significance, direct observation of multipolar order remains challenging.
Experimental evidence has so far relied on advanced techniques such as neutron scattering with large momentum transfer or high magnetic field, resonant x-ray scattering, and muon spin resonance~\cite{Kuwahara2007-xk,Portnichenko2020neutron,Hirota2000-pr,Nakao2001-hs,Takagiwa2002-df,Maharaj2020-fq}.
Recent studies propose the use of cross-correlated responses and nonlinear transport~\cite{Patri2019-ug,Kim2020-mp,Koizumi2023-ex,Grimmer2017-zh,Paramekanti2020-gt,Sorensen2021-om,Oike2024-rj,Bhowal2024-zy,Sorn2024-st,Ye2024-qv,Fang2024-ed}.
These promising approaches, however, require multiple external fields or highly sensitive detection of weak nonlinear signals and may suffer from partial signal compensation due to the domain states.
Thus, tabletop methodology is desirable for realizing direct and optical detection of multipolar order.

We propose dual-circular Raman optical activity as an all-optical probe of multipolar order.
Raman optical activity (ROA) represents the difference in Raman scattering intensities concerning circularly polarized light and is associated with chirality, magnetization, and ferroaxial order~\cite{Koshizuka1980-zz,Sirenko1997-oy,Kossacki2012-ft,Huang2020-hq,Cenker2021-bd,Liu2023-vc,Ishito2022-fr,Oishi2024-dy,Lacinska2022-ti,Yang2022-dl,Liu2023-qm,Kusuno_experiment}.
The dual-circular ROA denotes the dichroism related to the circularly polarized incident and scattered lights in the Raman scattering and consists of the two types, that is, cross-circular and parallel-circular ROA~\cite{Nafie1989-hj,Hecht1990-kp,Hecht1991-sw,Che1991-vn,Vargek1997-rv}.
We define $I_{\bm{e}_\text{i} \bm{e}_\text{f}} (\delta \omega, \omega)$ as the Raman scattering intensity for the incident and scattered light whose photon energy and polarization state are $\left(\omega, \bm{e}_\text{i}\right)$ and $\left(\omega + \delta  \omega, \bm{e}_\text{f}\right)$, respectively. 
The cross-circular ROA indicates the difference between $I_\text{LR}$ and $I_\text{RL}$ (L and R mean the left-handed and right-handed circular lights, respectively).
Similarly, the parallel-circular ROA is defined by the difference between $I_\text{LL}$ and $I_\text{RR}$.
These dichroisms indicate a preferred change of the circular polarization of photons in the Raman scattering event.

We found that the dual-circular ROA arises from the axial octupolar order, which is a representative example of multipolar order [Fig.~\ref{Fig_1}(a)], and provides a direct and symmetry-sensitive signature of their order parameters.
In contrast to conventional probes, which rely on the high-momentum-transfer or external fields to break multipolar symmetry, our analysis demonstrates that the cross-circular ROA arises under spatially uniform photoelectric fields and can therefore be of experimentally measurable magnitude.
The quantitative significance is evidenced by the first-principles calculations of the cross-circular ROA of pyrite, a prototypical material hosting the axial octupolar anisotropy.

Furthermore, the phonon excitations relevant to the proposed cross-circular ROA exhibit a characteristic three-dimensional motion, which can be viewed as a three-dimensional realization of chiral phonons~\cite{Zhang2014chiralphonon,Zhang2015-vy,Juraschek2017-jx,Ishito2022-fr,Zhang2022chiralphononTe,Hart2025-yc,sutcliffe2025pseudochiralphononsplittingoctupolar,Juraschek2025-ja,Zhang2025chiralphononReview}.
We found that the \textit{multipolar phonon} plays the vital role in the cross-circular ROA of axial octupolar materials.
Our work not only provides a convenient optical probe of the multipolar orders, which have been regarded as `hidden' due to the scarcity of the external stimuli coupled directly to them, but also implies the intriguing connection between the multipole physics and chiral phonon.

\noindent \textit{Results---}
We begin by considering the symmetry condition for dual-circular ROA.
The following symmetry analysis focuses on the cross-circular ROA defined as $U_\text{CC} = I_\text{LR} - I_\text{RL} $, although a similar discussion applies to the parallel-circular ROA defined by $U_\text{PC} = I_\text{LL} - I_\text{RR} $.
Supposing the backscattering geometry with the normal incidence of light along a $n$-fold ($n>2$) rotation axis, which eliminates spurious birefringence effect~\cite{Porto1966-wn,Hoffman1994-ik,Zhang2017-sh}, we obtain the following relation imposed by the out-of-plane mirror operation $m_\perp$ whose mirror plane contains the light beam axis:
                \begin{equation}
                U_\text{CC} (\delta \omega, \omega, \sca ) = - U_\text{CC} (\delta \omega, \omega,  m_\perp \sca).
                \end{equation}
Since the mirror operation interchanges the circular polarization of light, the sign of $U_\text{CC}$ is flipped.

The dependence on physical quantities, such as the crystal structure and order parameter, is denoted by $\sca$.
If $m_\perp \sca = \sca$ holds, cross-circular ROA is forbidden because $U_\text{CC} (\delta \omega, \omega, \sca) = 0$.
The symmetry for $m_\perp$ is broken by axial dipoles or chirality, which respectively preserve and break space-inversion symmetry~\cite{watanabe_ROA_symmetry}.
For instance, the magnetization $\bm{M}$ parallel to the incident light is inverted under the $m_\perp$ operation: $m_\perp \bm{M} = - \bm{M}$.

In the following, we consider the octupolar order belonging to the irreducible representation $A_{2g}^\tau$ of the cubic point group $m\bar{3}m$ where $\tau = \pm$ represents the time-reversal (\T{}) parity.
In context of the multipolar degrees of freedom in solids~\cite{Watanabe2018-do,Hayami2018-bh}, $A_{2g}^+$ and $A_{2g}^-$ orders correspond to the $xyz$-type axial and magnetic octupolar orders [Fig.~\ref{Fig_1}(a)].
This type of octupolar anisotropy neither breaks the cubic symmetry nor realizes the chirality, but violates the symmetry regarding the mirror operation $m_\perp$ perpendicular to $\left\{111\right\}$ and crystallographically-equivalent planes.
Conversely, the $m_\perp$ operation inverts the polarity of the axial octupolar order [Fig.~\ref{Fig_1}(b)].
As a result, the cross-circular ROA can occur when the light is incident on these planes.
From the perspective of crystallographic groups, taking the high-symmetric cubic point group $m\bar{3}m1'$ for the para-state, the ordered states characterized by $m\bar{3}1'$ (in the $A_{2g}^+$ case) and $m\bar{3}m'$ (in the $A_{2g}^-$ case) exhibit cross-circular ROA distinct from those observed in axial dipolar and chiral materials~\cite{Supplment}.

\begin{figure*}[htbp]
        \centering
        \includegraphics[width=0.75\linewidth,clip]{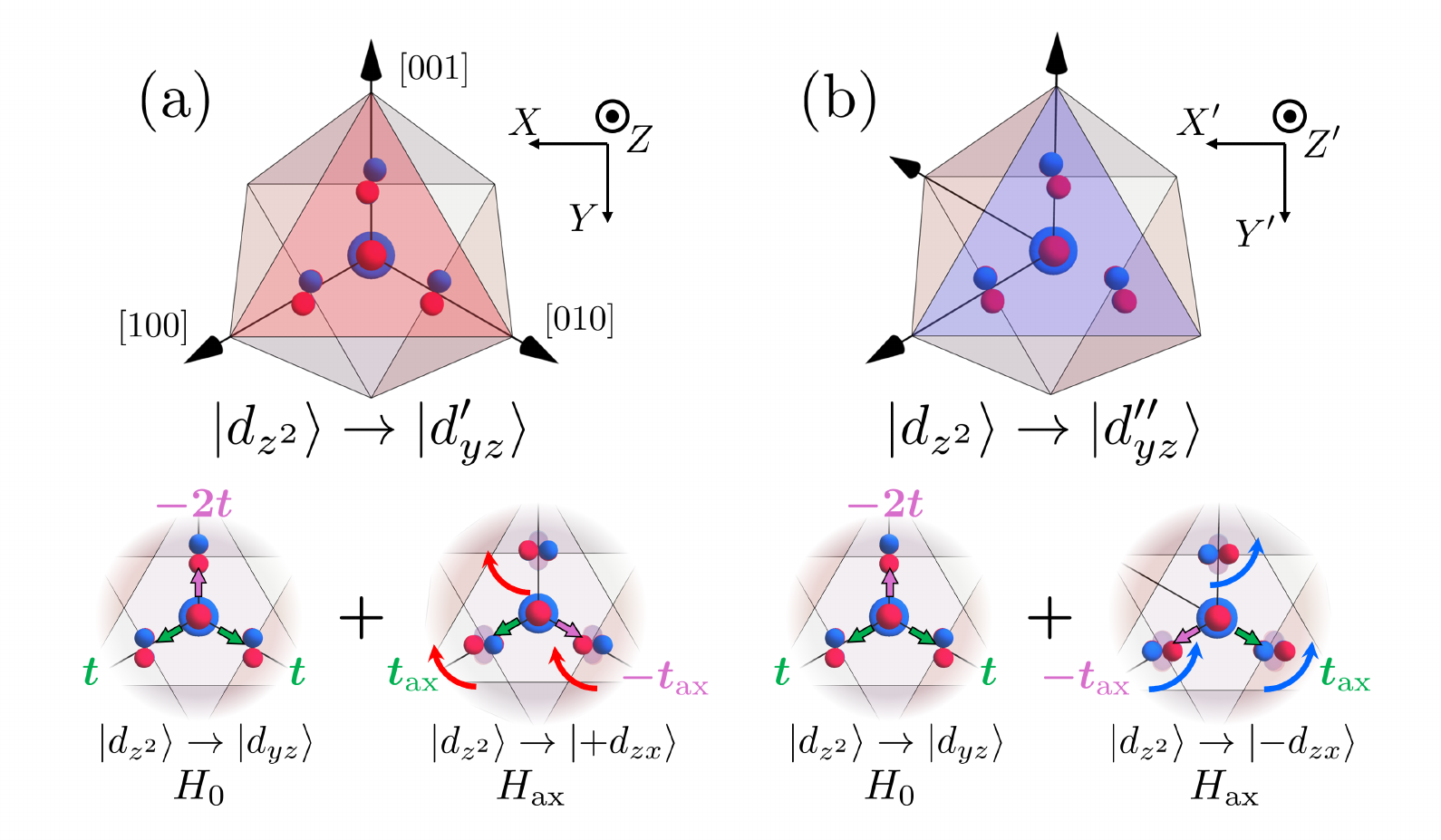}
        \caption{
                Hopping from $d_{z^2}$ to the perturbed $d_{yz}$ state, defined in the coordinate system with $X = [1\bar{1}0]$ and $Z = [111]$ in (a) and with $X' = [110]$ and $Z' = [\bar{1}11]$ in (b).
                The hopping process is decomposed into $d_{z^2} \to d_{yz}$ and $d_{z^2} \to \pm d_{zx}$, which result from the original ($H_0$) and axial octupolar ($H_\text{ax}$) terms, respectively.
                The hybridized $\ket{d_{zx}}$ state stems from the $\pm \pi/2$ rotation of the original $\ket{d_{yz}}$ state mediated by the face-dependent axial dipoles.    
        }
        \label{Fig_octupolar_hopping_rotation}
        \end{figure*}

Next, we demonstrate cross-circular ROA through a microscopic analysis.
The Raman scattering intensity is given by the semiclassical formula~\cite{Loudon2001-zi}, such as 
                \begin{equation}
                I_\text{LR} (\delta \omega, \omega) = \left| \bm{e}_\text{R}^\dagger \hat{\alpha} (\delta \omega, \omega) \bm{e}_\text{L} \right|^2,
                \end{equation}
where the right circularly polarized incident light and left circularly polarized scattered light are given by the polarization vectors $\bm{e}_\text{R} = (1,i,0)/\sqrt{2}$ and $\bm{e}_\text{L} = (1,-i,0)/\sqrt{2}$, respectively.
The Raman tensor $\hat{\alpha}$ for an elementary excitation is composed of the excitation $\Phi$ and the nonlinear susceptibility $\hat{\chi}^\Phi$, expressed as $\hat{\alpha} \equiv \Phi (\delta \omega) \hat{\chi}^\Phi (\delta \omega, \omega)$.
The nonlinear susceptibility is defined by the relation
                \begin{equation}
                P_a (\omega + \delta \omega) = \sum_b \chi_{ab}^\Phi (\delta \omega, \omega) \Phi (\delta \omega )E_b (\omega),
                \label{formula_nonlinear_susceptibility}
                \end{equation}
where $a,b = x,y,z$.
The formula describes the correction to the electric susceptibility induced by $\Phi$ with a frequency shift $\delta \omega$.

The cross-circular ROA originates from the excitations $\Phi_{1}$ and $\Phi_{2}$, which belong to the complex irreducible representations $E_g^1$ and $E_g^2$ of $m\bar{3}$.
The symmetry-adapted form of $\chi_{ab}^\Phi$ for each excitation is given by 
                \begin{equation}
                        \hat{\chi}^{(1)} = \chi_1 \text{diag} \left( 1, \xi^2,\xi \right),
                        \hat{\chi}^{(2)} = \chi_2 \text{diag} \left( 1, \xi,\xi^2 \right),
                        \label{Eg_chi_symmetry_adapted_formula}
                \end{equation}
where $\xi = \text{exp} (2\pi i /3)$~\footnote{
 Note that $\hat{\chi}^{(1)}$ belongs to the $E_g^2$ irreducible representation, while $\hat{\chi}^{(2)}$ to the $E_g^1$.
}.
The Raman tensors represent dynamics characterized by sequential stretching and contraction, occurring in the order of [100]-[001]-[010] directions and [100]-[010]-[001] directions~\cite{Supplment}.
Intriguingly, those dynamics represent the three-dimensional extension of the chiral phonons~\cite{Zhang2014chiralphonon,Ishito2022-fr,Hart2025-yc}.
More explicitly, while the eigenmode of the chiral phonon is defined in a plane, the $E_g^{1,2}$-mode dynamics cannot be reduced to that in a plane or line, and thus the corresponding displacement may be called a multipolar phonon.

The cross-circular Raman scattering intensities for the light incidence along $[111]$ are obtained as
                \begin{align}
                I_{\text{LR}}^{[111]}(\delta \omega,\omega) &=  \left|  \chi_1 (\delta \omega, \omega) \Phi_1 (\delta \omega) \right|^2, \label{LR_scattering_amplitude}\\
                I_{\text{RL}}^{[111]}(\delta \omega,\omega) &=  \left|  \chi_2 (\delta \omega, \omega) \Phi_2 (\delta \omega) \right|^2,
                \label{RL_scattering_amplitude}
                \end{align}
and hence we obtain the corresponding ROA as
            \begin{equation}
                U_\text{CC}^{[111]} =  \left|  \chi_1 (\delta \omega, \omega) \Phi_1 (\delta \omega) \right|^2 - \left|  \chi_2 (\delta \omega, \omega) \Phi_2 (\delta \omega) \right|^2,
                \label{CCROA_111_face}
            \end{equation}
Thus, the cross-circular ROA arises from differences in the magnitudes of $\hat{\chi}^{(1,2)}$ and $\Phi_{1,2}$.
Furthermore, the octupolar anisotropy and the multipolar phonon give rise to cross-circular ROA whose sign alternates depending on the incident plane.
For instance, $\chi_1$ and $\Phi_1$ contribute to $I_\text{LR}$ for the light incidence along $[111]$, while they contribute to $I_\text{RL}$ for the incidence along $[\bar{1}11]$.
For the latter case, the ROA is given by
                \begin{equation}
                        U_\text{CC}^{[\bar{1}11]} 
                                =  \left|  \chi_2 (\delta \omega, \omega) \Phi_2 (\delta \omega) \right|^2 - \left|  \chi_1 (\delta \omega, \omega) \Phi_1 (\delta \omega) \right|^2,
                \end{equation}
equal to $-U_\text{CC}^{[111]}$.
As illustrated in Fig.~\ref{Fig_1}(c,d), the alternating signs of ROA reflect the octupolar configuration of the axial vectors penetrating the faces of the octahedron.

Let us now consider the \T{}-even case, that is $A_{2g}^+$ octupolar system.
In phonon-mediated Raman scattering, the degenerate multipolar phonon modes $\Phi_{1,2} (\delta \omega)$ are equally excited in equilibrium because of the \T{} symmetry, such that $\left| \Phi_1 (\delta \omega) \right| = \left| \Phi_2 (\delta \omega) \right|$.
Consequently, the cross-circular ROA for the $\left\{111\right\}$ face, $U_{\text{CC}}^{[111]} (\delta \omega,\omega) $, occurs by the imbalance between the magnitudes of nonlinear susceptibilities $\left|\chi_{1,2} (\delta \omega, \omega) \right|^2$.

We demonstrate the multipolar cross-circular ROA by considering the full set of spinless $d$-orbital fermions residing on a primitive cubic lattice.
The tight-binding Hamiltonian is given by $H = \sum_\bk H_0 (\bk) + H_\text{ax} (\bk)$.
$H_0 (\bk)$ shows the cubic symmetry ($m\bar{3}m$) and is parametrized using the Slater-Koster method.
The axial multipolar term $H_\text{ax} (\bk)$, governed by the single parameter $t_\text{ax}$, is derived from the multipole-based analysis~\cite{harrison2012electronic,Kusunose2023-ft}.
Coupling to the external fields, such as the electric field and phonon excitations, is also included to evaluate the nonlinear susceptibility.
The electromagnetic perturbation is treated via the minimal coupling between the momentum and vector potential as $\bm{p} \to \bm{p}-\bm{A}$.
We note that the electromagnetic fields are treated with the electric-dipole approximation, and hence no spatial dispersion of light is taken into account.
The multipolar phonon excitations $\Phi_{1,2}$ are introduced through the perturbative Hamiltonian
                \begin{equation}
                \delta H_\Phi = \sum_\bk \Phi_{1(2)} \, \bm{c}_\bk^\dagger \text{diag} (1,\xi^{\pm 1},\xi^{\mp 1}) \bm{c}_\bk,
                \end{equation}
where the creation operators of $t_{2g} $ manifold denotes $\bm{c}_\bk^\dagger = \left( d_{yz}^\dagger,d_{zx}^\dagger,d_{xy}^\dagger \right)$~\footnote{Although the operator $\delta H_\Phi$ is non-hermitian, the present calculation is linear in $\Phi_{1,2}$ and therefore does not suffer from the non-hermitian property due to the superposition.}.
Further details of the model are provided in Ref.~\cite{Supplment}.

\begin{figure}[htbp]
                \centering
                \includegraphics[width=0.75\linewidth,clip]{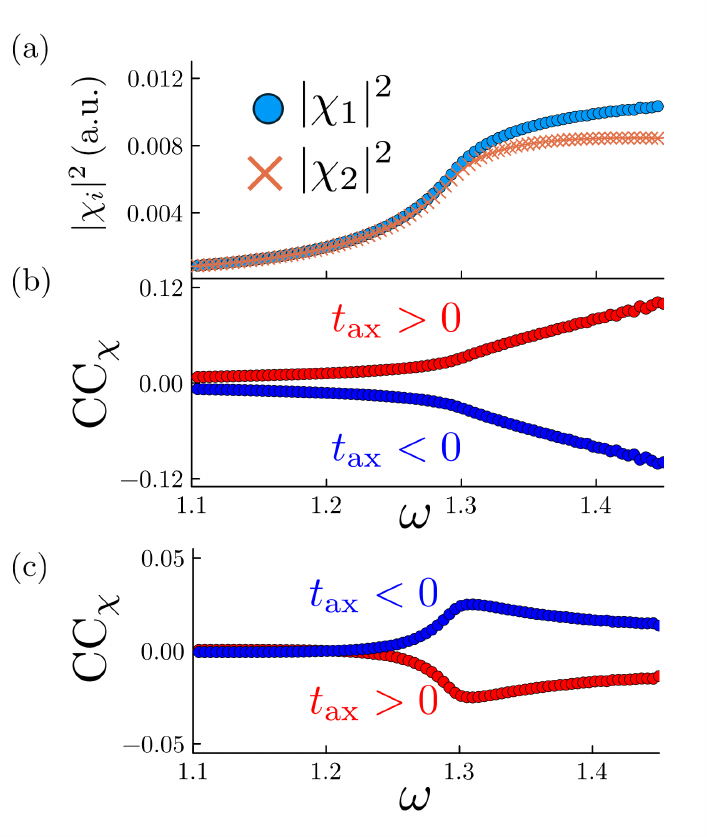}
                \caption{
                        Dependence on the incident light frequency $\omega$ of (a) nonlinear susceptibilities $|\chi_1|^2$ and $|\chi_2|^2$ for the $A_{2g}^+$ octupolar system and (b) indicator for cross-circular ROA CC$_\chi \equiv \left(   | \chi_1|^2 - | \chi_2|^2 \right) / \left(   | \chi_1|^2 + | \chi_2|^2 \right)$.
                        The two polarities of the $A_{2g}^+$ octupolar order are considered in (b).
                        (c) $\omega$ dependence of CC$_\chi$ for the $A_{2g}^-$ octupolar systems with opposite polarities.
                        $\delta \omega = 0.1$ and $t_\text{ax} = \pm 0.1$ are used.
                        }
                \label{Fig_roa_freq}
                \end{figure}

The nonlinear susceptibilities $\chi_{1,2}$ are microscopically computed by the established perturbative approach~\cite{Supplment}.
The numerical calculations confirm that the obtained susceptibilities are adapted to each irreducible representation as $\chi_1 \equiv \chi_{xx}^{(1)} = \xi \chi_{yy}^{(1)}  = \xi^{-1} \chi_{zz}^{(1)}$ and $\chi_2 \equiv  \chi_{xx}^{(2)} = \xi^{-1} \chi_{yy}^{(2)}  = \xi \chi_{zz}^{(2)}$, consistent with Eq.~\eqref{Eg_chi_symmetry_adapted_formula}.
Figure~\ref{Fig_roa_freq}(a) presents the dependence of $| \chi_2|^2$ and $| \chi_1|^2$ on the incident light frequency $\omega$.
It is evident that the two susceptibilities differ significantly from each other, and the difference $| \chi_1|^2 - | \chi_2|^2$ reverses sign when the octupolar parameter $t_\text{ax}$ is inverted.
In Fig~\ref{Fig_roa_freq}(b), we show the normalized difference CC$_\chi \equiv \left(   | \chi_1|^2 - | \chi_2|^2 \right) / \left(   | \chi_1|^2 + | \chi_2|^2 \right)$.
The dichroism is particularly pronounced in the presence of the resonant particle-hole excitations ($\omega \gtrsim 1.2$), similar to the ROA observed in the ferroaxial~\cite{Kusuno_experiment} and antisymmetric Raman scattering~\cite{Koningstein1968-ci,Mortensen1968-gn}.

Let us corroborate the physical origin of the cross-circular ROA.
The $E_g^{1}$ and $E_g^{2}$ multipolar phonons show the circular motion in the $\left\{111\right\}$ and its equivalent planes, thereby coupled to the electron's orbital rotation induced by the axial octupolar order.
To elucidate the orbital rotation, we analyze the hopping symmetry between the $d_{z^2},d_{yz}$ orbitals with the cartesian coordinates defined by the $X = [1\bar{1}0]$ and $Z=[111]$ [Fig.~\ref{Fig_octupolar_hopping_rotation}(b)].
The nearest-neighbor hopping parameter for the non-octupolar part is given by $t_\alpha (\bk) \equiv \Braket{d_{yz} | H_0 (\bk) | d_{z^2}}$, while that for the axial octupolar part is by $t'_\alpha (\bk) \equiv \Braket{d_{yz} | H_\text{ax} (\bk) | d_{z^2}} $ [Fig.~\ref{Fig_octupolar_hopping_rotation}(a)].
The obtained multipolar hopping is related to the nearest-neighbor hopping concerning the $d_{zx}$ orbital as $t'_\alpha (\bk) \propto \text{sign}(t_\text{ax}) \cdot t_\beta (\bk)$ ($t_\beta (\bk) \equiv \Braket{d_{zx} | H_0 (\bk) | d_{z^2}}$)~\cite{Supplment}.
The similarity between $t'_\alpha$ and $t_\beta$ is attributed to the perturbed state $\ket{d'_{yz}} = \ket{d_{yz}} + \delta \ket{d_{zx}}$ ($\delta \propto t_\text{ax}$) by which the hopping process $d_{z^2} \to d_{zx}$ is admixed with the $d_{z^2} \to d_{yz}$ hopping.
The hybridization between $\ket{d_{yz}}$ and $\ket{d_{zx}}$, denoting the orbital rotation along the $Z$ axis, is consistent with nature of the axial dipole pointing to $Z=[111]$, as the axial dipole relate the quantities in a transversal manner~\cite{Hayami2022-yq}.
The orbital rotation leads to the cross-circular ROA through its coupling to the circular motion of the $E_g^{1,2}$ phonons.
This physical picture has been supported by the analytical calculations for a ferroaxial system~\cite{Kusuno_experiment,Oiwa2022-he}.

Owing to the octupolar order, the axial dipoles and the associated orbital rotations alternate between the faces of the octahedron.
For example, in the coordinate system defined by $X' = [110]$ and $Z' = [\bar{1}11]$, the $\ket{d_{yz}}$ state undergoes the opposite orbital rotation as described by $\ket{d''_{yz}} = \ket{d_{yz}} - \delta \ket{d_{zx}}$ [Fig.~\ref{Fig_octupolar_hopping_rotation}(b)].
Consequently, the cross-circular ROA exhibits its incident-plane dependence in agreement with the symmetry analysis.

Cross-circular ROA is similarly realized in the $A_{2g}^-$ ($\theta$-odd) octupolar order.
The magnetoaxial octupolar term of $H_\text{ax}$ is expressed by the orbital-current order which reduces the point group symmetry as $m\bar{3}m1' \to m\bar{3}m'$~\cite{Supplment}.
As in the \T{}-even case, supposing the phonon excitation $\Phi_{1,2}$ in Eq.~\eqref{formula_nonlinear_susceptibility}, we observe that the nonzero CC$_\chi$ shows the sign reversal under inversion of the $A_{2g}^- $ multipolar order [Fig.\ref{Fig_roa_freq}(c)].

Moreover, in addition to the imbalance in nonlinear susceptibilities, the $A_{2g}^-$ order lifts the degeneracy of the $\Phi_{1,2} (\delta \omega)$ multipolar phonons, giving rise to the different $\delta \omega$ dependence [Eq.~\eqref{CCROA_111_face}].
The \T{}-odd nature allows for the energy gain phenomenologically given by
                \begin{equation}
                \delta E_\Phi (\delta \omega) =   \eta_\text{ax}\,  \delta \omega \left[ \left| \Phi_1 (\delta \omega )\right|^2 - \left| \Phi_2 (\delta\omega ) \right|^2  \right],
                \label{a2g_phonon_lifting}
                \end{equation}
where $\eta_\text{ax}$ ($\propto t_\text{ax}$) stands for the coupling between the $A_{2g}^-$ order and the multipolar motion of phonons expressed as $\bm{\Phi} \times \partial_t \bm{\Phi}$ for $\bm{\Phi} = \left( \text{Re}{\Phi_1}, \text{Im}{\Phi_1} \right) = \left( \text{Re}{\Phi_2}, - \text{Im}{\Phi_2} \right)$.
Thus, the resonant frequencies of $\Phi_{1,2} (\delta \omega)$ are slightly shifted in a staggered manner as $\delta \omega_{\pm} = \Omega_0 \pm \delta \Omega$ where $\delta \Omega \propto \eta_\text{ax}$ as in the case of electrons' energy levels~\cite{shibata2025Octupolar}.
As a result, the Raman tensor $\hat{\alpha} (\delta \omega_+, \omega)$ is predominantly determined by $\Phi_1 (\delta \omega_+)$, while $\hat{\alpha} (\delta \omega_-, \omega)$ is dominated by $\Phi_2 (\delta \omega_-)$, pointing to the cross-circular ROA for $\delta \omega = \delta \omega_\pm $ without compensation of the contributions of $\Phi_1$ and $\Phi_2$.
In contrast, for the \T{}-even case, the modes $\Phi_{1,2} (\delta \omega)$ remain degenerate due to no \T{}-odd coupling $\eta_\text{ax} = 0$.

Intriguingly, the Zeeman-like splitting of Eq.~\eqref{a2g_phonon_lifting} leads to the contrasting Stokes and anti-Stokes signals, which are the Raman scattering processes involving the creation and annihilation of phonons.
The mode $\Phi_2 (-\delta \omega_+)$ plays a dominant role in $\hat{\alpha} (-\delta \omega_+, \omega)$, while $\Phi_1 (-\delta \omega_-)$ does in $\hat{\alpha} (-\delta \omega_-, \omega)$ as evident from the prefactor $\delta \omega$ of Eq.~\eqref{a2g_phonon_lifting}.
Thus, the cross-circular ROA $U_\text{CC} (\pm |\delta \omega|, \omega)$ stemming from the $A_{2g}^-$ octupolar order shows the antisymmetry between the Stokes ($- |\delta \omega| <0$) and anti-Stokes ($+ |\delta \omega| >0$) signals, since each signal is primarily governed by either of the $\Phi_1$ or $\Phi_2$ excitation~\cite{watanabe_ROA_symmetry}.
The antisymmetry is similarly found in the $\theta$-odd axial dipolar order (\textit{e.g.,} ferromagnetic order)~\cite{Cenker2021-bd}.

On the other hand, the $A_{2g}^+$ octupolar state shows the Stokes and anti-Stokes cross-circular scattering intensities with the same signs, because the nonlinear susceptibilities do not show significant change between $|\chi_{i} (\delta \omega, \omega)|$ and $|\chi_{i} (- \delta \omega, \omega)|$ for $i=1,2$ under the condition $|\delta \omega| \ll |\omega|$~\cite{Supplment}.
It follows that the careful comparison of Stokes and anti-Stokes signals provides a means to identify the time-reversal parity of the multipolar order~\cite{watanabe_ROA_symmetry,Barron1976-hs,Loudon1978-ud,Barron1985-ia,Hecht1993-ag}.

\begin{figure}[htbp]
                \centering
                \includegraphics[width=0.90\linewidth,clip]{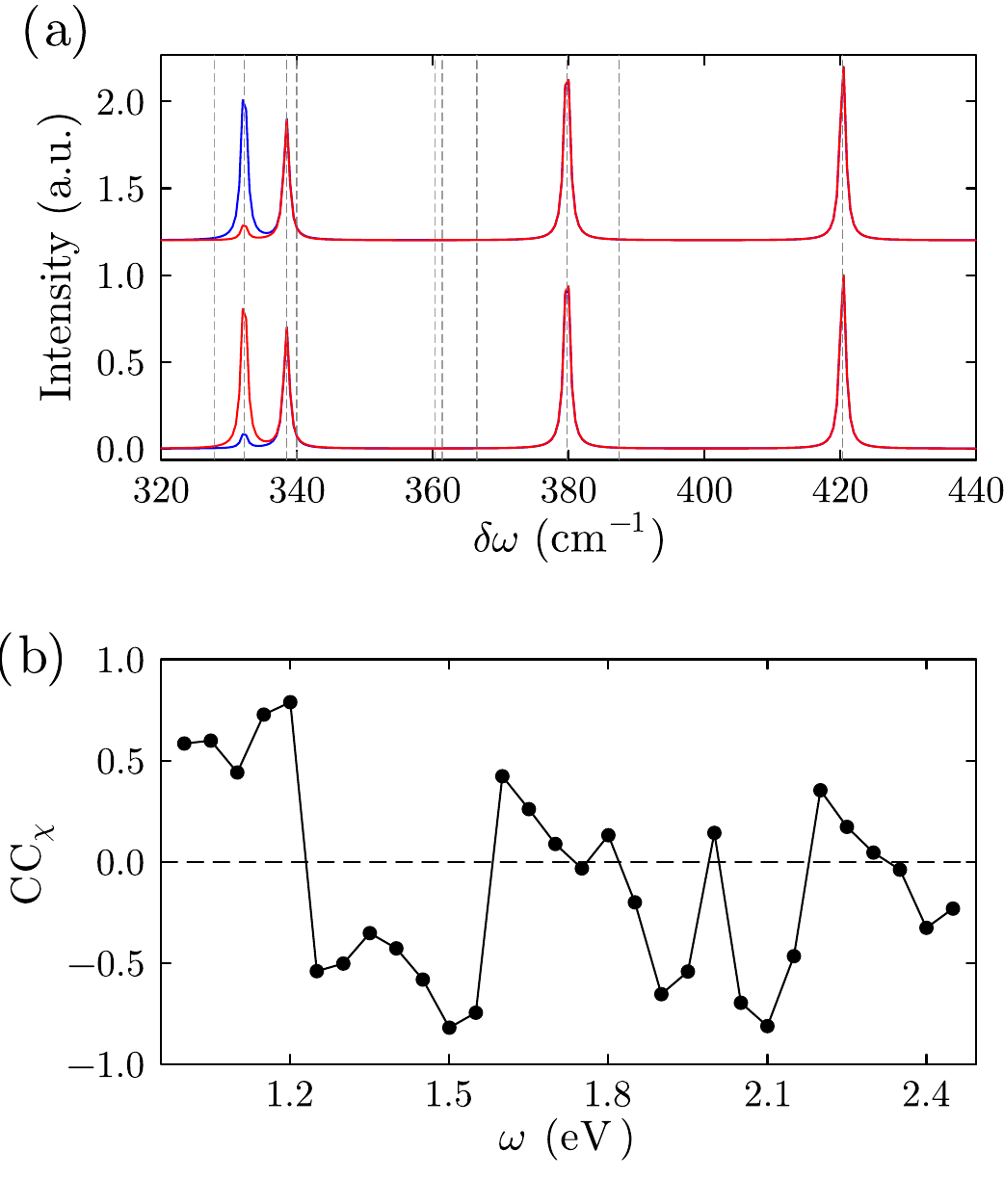}
                \caption{
                        (a) Raman spectrum of pyrite for $\omega = 1.5$~(eV).
                        The dotted vertical lines denote the phonon energies at the Brillouin zone center.
                        The incidence directions are $[111]$ and $[\bar{1}11]$ for the upper and lower panels.
                        In these configurations, the $E_g^{1,2}$ [$\delta \omega \sim 322$~(cm$^{-1}$)] and $T_g$ modes [$\delta \omega \sim 340,~380,~420$~(cm$^{-1}$)] are Raman active.
                        The $E_g^{1,2}$ peak shows the cross-circular ROA with opposite signs between the two configurations, while the $T_g$ modes do not show ROA.
                        The red and blue plots are for $I_\text{LR}$ and $I_\text{RL}$, respectively.
                        (b) ROA indicator CC$_\chi$ and its dependence on the light frequency $\omega$ for the $[111]$ incidence.
                        }
                \label{Fig_fes2}
                \end{figure}

Finally, we present the first-principles analysis of the cross-circular ROA of pyrite (FeS$_2$), demonstrating that the multipolar ROA attains a substantial magnitude.
Its crystal structure is labeled by the space group $Pa\bar{3}$ (No.~205) whose point group $m\bar{3}1'$ is of our interest.
In Fig.~\ref{Fig_fes2}, we plot the Raman spectrum and the normalized ROA $\text{CC}_{\chi} = \text{CC}_{\chi} (\omega)$.
In agreement with the preceding discussions, the cross-circular ROA is observed in the $E_g^{1,2}$ modes at $\delta \omega = 332$ (cm$^{-1}$).
The facet-dependent ROA is also consistent with our predictions [Fig.~\ref{Fig_fes2}(a)].
Importantly, the ROA is sizable as much as $\text{CC}_\chi$ amounting to several ten percents [Fig.~\ref{Fig_fes2}(a)].
The quantitative significance is promising for the detection of the multipolar order and is consistent with the experimental result~\cite{Suganuma-experiment}.

\noindent\textit{Discussions---}
We proposed the dual circular ROA originating from the axial octupolar order.
By integrating the symmetry analysis and microscopic calculations, we have demonstrated the emergence of cross-circular ROA in the $A_{2g}^+$ (\T{}-even) and $A_{2g}^-$ (\T{}-odd) axial multipolar phases.
This activity enables all-optical measurement of multipolar orders and further suggests its potential as a tabletop probe of multipolar order, and indicates the close relation with chiral phonon.
The close relation further implies that, by combing the circular-light driving of multipolar phonons~\cite{He2024-ie,Hart2025-yc}, one can realize the all-optical write and read protocol for the axial octupolar order.

Though the quantitative significance has been demonstrated by our first-principles calculations, further material-specific investigations are highly desirable.
For instance, the \T{}-odd type is expected in the all-in-all-out spin state of pyrochlore magnet, noncoplanar spin state of rare-earth galium (aluminum) garnets, and the $\theta$-odd octupolar state of spin-orbit-coupled systems~\cite{Onimaru2011-ru,Sakai2011-la,Gallego2016-yz,Paramekanti2020-gt,Voleti2020-hq,Maharaj2020-fq,Voleti2023-cz}.
For the \T{}-even case, to the best of our knowledge, no experimental report has been made on the axial octupolar ordering resulting in the symmetry reduction from $m\bar{3}m1' \to m\bar{3}1'$.
The phase transition of interest might be realized in the rare-earth oxides ${R_2}$O$_3$ ($R$~: rare earth) or double perovskites, where the crystal structures can be tuned via pressure or chemical substitution~\cite{Jiang2013-hk,Saroj2018-uy}.
Further exploration of the octupolar dual-circular ROA in real materials is an important direction for future research.

\noindent
\textit{note added---}
We thank A. Paramekanti for bringing their work on arXiv~\cite{sutcliffe2025pseudochiralphononsplittingoctupolar} to our attention.
They clarified that the $E_g^1$ ($E_g^2$) mode shows the pseudo angular momentum as in the case of chiral phonons and is split by the $\theta$-odd octupolar order, related to Eqs.~\eqref{Eg_chi_symmetry_adapted_formula} and~\eqref{a2g_phonon_lifting}.

\section*{acknowledgement}
We thank Gakuto Kusuno, Takuya Nomoto, Takuya Satoh, and Yuki Suganuma for fruitful comments and discussions.
H.W. thanks Atsushi Fujimori and Jun Kobayashi for helpful comments.
R.O. was supported by Special Postdoctoral Researcher Program at RIKEN.
This work is supported by Grant-in-Aid for Scientific Research from JSPS KAKENHI Grant
No.~JP23K13058 (H.W.),
No.~JP24K00581 (H.W.),
No.~JP25H02115 (H.W.),
No.~JP25K23349 (H.M.),
No.~JP21H04990 (R.A.),
No.~JP25H01246 (R.A.),
No.~JP25H01252 (R.A.),
JST-CREST No.~JPMJCR23O4(R.A.),
JST-ASPIRE No.~JPMJAP2317 (R.A.),
JST-Mirai No.~JPMJMI20A1 (R.A.),
and RIKEN TRIP initiative (RIKEN Quantum, Advanced General Intelligence for Science Program, Many-body Electron Systems).
H.W. was also supported by JSR Corporation via JSR-UTokyo Collaboration Hub, CURIE.

\section*{End Matter}

Parallel-circular ROA, which is defined by $I_\text{LL} - I_\text{RR}$, can be computed similarly to the case of cross-circular ROa, but it is anticipated to be weak because relevant Raman scattering does not occur under the spatially uniform electric fields.

The parallel-circular ROA arises from the interference of the totally symmetric and antisymmetric Raman scatterings~\cite{Nafie1989-hj}.
In the $A_{2g}^+$ case, for example, the excitation $\Phi_\text{PC}$ responsible for the parallel-circular Raman scattering belongs to the $A_g$ representation of $m\bar{3}$.
However, the $A_g$-adapted Raman tensor contains only the totally-symmetric part ($\hat{\alpha} \propto 1_3 $) under the electric-dipole approximation.
The lowest-order antisymmetric contribution to parallel-circular ROA is invoked by the spatial variation of the electromagnetic fields, like octupole transitions~\cite{Supplment}.
Taking into account the electric octupolar fields such as $ \partial_x \partial_y  E_x $, we obtain the $A_g$-adapted Raman tensor as
                \begin{equation}
                \hat{\alpha} = \Phi_{A_g}
                \begin{pmatrix}
                        \alpha & i \beta q_x q_y &- i\beta q_z q_x\\
                        - i\beta q_x q_y & \alpha&i\beta q_y q_z\\
                        i\beta q_z q_x  & -i\beta q_y q_z  & \alpha
                \end{pmatrix},
                \end{equation}
where $\bm{q}$ is the wavevector of the incident light.
$\alpha$ and $\beta$ denote the nonlinear susceptibilities without and with the account for electric-octupole transitions, respectively. 
The obtained Raman tensor leads to parallel-circular ROA whose sign depends on both the polarity of the multipolar order and the incident plane of light, the latter of which comes from the $\bm{q}$ dependence.
The resultant ROA may be hard to detect due to the small momentum transfer from photons to electrons, being in stark contrast to cross-circular ROA occurring within the electric-dipole approximation.

\end{document}